\documentstyle[prl,aps,epsf]{revtex}
\newcommand{\Pint}{-\hspace{-1.05em}\int}
\newcommand{\E}{{\cal{E}}}

\newcommand{\I}{{\rm i}}
\renewcommand{\d}{{\rm d}}  

\newcommand{\be}{\begin{equation}}      
\newcommand{\ee}{\end{equation}}
\newcommand{\bea}{\begin{eqnarray}}      
\newcommand{\eea}{\end{eqnarray}}
\def\J#1#2#3#4{{#1} {\bf #2}, #3 (#4)}
\def\PRL{Phys. Rev. Lett.}
\def\PRD{Phys. Rev. D}
\def\PR{Phys. Rev.}
\def\PTP{Prog. Theor. Phys.}
\def\AAR{Astron. Astrophys. Rev.}
\def\AA{Astron. Astrophys.}
\def\AJ{Ap. J.}
\def\AL{Astron. Lett.}
\def\JMP{J. Math. Phys.}
\def\CQG{Class. Quantum Grav.}
\def\ZP{Z. Phys.}
\def\PLA{Phys. Lett. A}

\begin{document}
\preprint{}
\draft
\title{Exact Solution for the Exterior Field of a Rotating Neutron Star}

\author{Vladimir~S.~Manko$^\diamond$\thanks{E-mail: 
VladimirS.Manko@fis.cinvestav.mx}, 
Eckehard~W.~Mielke$^\$$\thanks{E-mail: ekke@xanum.uam.mx},
and Jos\'e~D.~Sanabria-G\'omez$^\diamond$\thanks{E-mail: 
Jose.Sanabria@fis.cinvestav.mx}
} 
\address{$^\diamond$ Departamento de F\'\i sica,\\ 
Centro de Investigaci\'on y de Estudios Avanzados del IPN,\\ 
A.P. 14-740, 07000 M\'exico D.F., Mexico\\
$^\$ $ Departamento de F\'\i sica,\\ Universidad Aut\'onoma
Metropolitana-Iztapalapa\\
A.P. 55-534, C.P. 09340 M\'exico D.F., Mexico}

\maketitle

\begin{abstract} A four-parameter class of exact asymptotically flat
solutions of the Einstein-Maxwell equations involving only rational
functions is presented.  It is able to describe the exterior field of
a slowly or rapidly rotating neutron star with poloidal magnetic
field.
\end{abstract}

\vspace{.15in}
\centerline{PACS numbers: 04.20Jb, 04.40N, 97.60L}

\vspace{2.5cm}


\newpage

\section{Introduction}

Observations of binary pulsars \cite{TAMT93} indicate that the
individual {\em neutron stars} (NS) in such systems have masses very
close to the Chandrasekhar limit of 1.4 $M_\odot$ of white dwarfs.
Theoretically, the issue of the maximum mass of a NS hings strongly
on the equation of state (EOS) and the particle interactions at the
high density of the center. Models of NS with strong kaon
condensation \cite{BB94} or even a quark nucleus (`strange star' 
\cite{GHL99}) can have at most 1.5 -- 2 $M_\odot$, which would leave
range for lower mass black holes. However, even modest {\em
differential rotation} \cite{BSS99} may easily increase the maximum
mass 2 $M_\odot$ of a nonrotating NS to above 3 $M_\odot$ for a
nascient NS in a transient phase of a supernova. Moreover, a
mass-quadrupole moment $Q$ is also important \cite{SS} for achieving
correspondence with numerical results \cite{C,LP99}. 
  
In order to model analytically the exterior field of a NS in the
framework of {\em general relativity} (GR), one needs an {\em exact}
asymptotically flat solution of the {\em Einstein-Maxwell equations}
(electrovac spacetimes) possessing at least four arbitrary physical
parameters which are the mass $M$, angular momentum $J$, magnetic
dipole $\mu$ and mass-quadrupole moment $Q$. The simplest solution,
besides, can be envisaged as axially symmetric, the magnetic field
sharing the symmetry of the mass and angular momentum distributions;
an additional reflection symmetry with respect to the equatorial
plane which is expected of the self-gravitating objects (see, e.g.,
\cite{MN} and references therein) must be also imposed. 

Our paper aims at presenting an exact solution which does satisfy
the above requirements and, what is most important, is a
mathematically very {\em simple} solution admitting a representation
exclusively in terms of the rational functions of spheroidal
coordinates (previous effort in this direction only led either to the
solution which had no reflection symmetry \cite{A-S} or to solutions
\cite{M-Z} which did not permit the rational function representation,
and consequently could not be written in a concise form).

\section{Four--parameter exact solution}

The reported solution has been constructed with the aid of
Sibgatullin's method \cite{S} according to which the complex
potentials $\E$ and $\Phi$ satisfying Ernst's equations \cite{E} 
are defined, for specified axis data $e(z):={\cal E}(z,\rho=0)$
and $f(z):=\Phi(z,\rho=0)$, by the integrals
\be
{\cal E}(z,\rho)=\frac{1}{\pi}\int_{-1}^1
\frac{e(\xi)\mu(\sigma)d\sigma}{\sqrt{1-\sigma^2}},
\quad
\Phi(z,\rho)=\frac{1}{\pi}\int_{-1}^1
\frac{f(\xi)\mu(\sigma)d\sigma}{\sqrt{1-\sigma^2}}\, .
\label{Ernst}
\ee
The unknown function $\mu(\sigma)$ is to be found from the singular
integral equation
\be
\Pint_{-1}^{1}\frac{\mu(\sigma)[e(\xi)+\tilde           
e(\eta)+2f(\xi)\tilde
f(\eta)]d\sigma}{(\sigma-\tau)\sqrt{1-\sigma^2}}=0
\ee
with the normalizing condition
\be
\int_{-1}^1\frac{\mu(\sigma)d\sigma}{\sqrt{1-\sigma^2}}=\pi,
\ee
where $\xi=z+\I\rho\sigma$, $\eta=z+\I\rho\tau$, $\rho$ and $z$ being
the Weyl-Papapetrou cylindrical coordinates and
$\sigma,\tau\in[-1,1]$; $\tilde e(\eta):=\overline{e(\bar\eta)}$,
$\tilde f(\eta):=\overline{f(\bar\eta)}$, and the overbar stands for
complex conjugation. 

In what follows, the axis data $e(z)$ and $f(z)$ are chosen in the
form
\bea
e(z)&=&\frac{(z-M-\I a)(z+\I b)+d-\delta -ab}
{(z+M-\I a)(z+\I b)+ d-\delta -ab}, \nonumber \\
f(z)&=&\frac{\I\mu}
{(z+M-\I a)(z+\I b)+ d-\delta -ab}, \nonumber \\
\delta &:=&\frac{\mu^2-M^2b^2}{M^2-(a-b)^2}, \qquad 
d:= \frac{1}{4}[M^2-(a-b)^2],
\eea
such that the algebraic equation
\be
e(z)+\bar e(z)+2f(z)\bar f(z)=0
\label{alge}
\ee
will have a pair of distinct roots of multiplicity two. This is a key
point for having the rational form of the final expressions for
${\E}(\rho,z)$ and $\Phi(\rho,z)$ after performing the
Riemann-Hilbert procedure of the analytic continuation of the
functions $e(z)$, $f(z)$ into the complex plane $(\rho,z)$. The
resulting expressions for the potentials ${\E}(\rho,z)$ and
$\Phi(\rho,z)$ obtained from Eqs.~(\ref{Ernst})-(\ref{alge}) are of
the following {\em polynomial} form:\footnote{All the formulas of
this paper have been checked with the aid of the Mathematica~3.0
computer program \cite{W}.}
\bea
\E&=&(A-2MB)/(A+2MB), \quad \Phi=2\I\mu C/(A+2MB),\nonumber\\
A&=&4[(k^2x^2-\delta  y^2)^2-d^2-\I k^3xy(a-b)(x^2-1)] \nonumber\\
&&-(1-y^2)[(a-b)(d-\delta )-M^2b][(a-b)(y^2+1)+4\I kxy], \nonumber\\
B&=&kx\{ 2k^2(x^2-1)+[b(a-b)+2\delta ](1-y^2)\} \nonumber\\
&&+\I y\{2k^2b(x^2-1)-[k^2(a-b)-M^2 b-2a\delta ](1-y^2)\},
\nonumber\\
C&=&2k^2y(x^2-1)+[2\delta  y-\I kx(a-b)](1-y^2)\, , 
\eea
where we have introduced the generalized spheroidal coordinates 
\bea
x&=&\frac{1}{2k}(r_+ +r_-) \quad {\rm and} \quad y=\frac1{2k}(r_+-r_-)\,,
\nonumber\\ 
r_\pm &:=&\sqrt{\rho^2+(z\pm k)^2}\, , \quad k:=\sqrt{d+\delta }\,.
\eea

The four arbitrary real parameters entering the solution are the
total mass $M$, total angular momentum per unit mass $a:=J/M$,
magnetic dipole moment $\mu$ and mass-quadrupole moment
\be
Q=-\frac{M}{4[M^2-(a-b)^2]}\left[ M^4+ 2M^2(a^2+b^2) -(3a+b)(a-b)^3 
-4\mu^2\right] \label{quadru}
\ee
of the source. 

The corresponding complete metric is given by the axisymmetric line
element\footnote{Throughout the paper, natural units are used in
which the gravitational constant and the velocity of light are equal
to unity.}
\be
\d s^2=-f(\d t-\omega\d\varphi)^2+
k^2f^{-1}\Bigl[e^{2\gamma}(x^2-y^2)
\Bigl(\frac{\d x^2}{x^2-1}+\frac{\d y^2}{1-y^2}\Bigr) 
+(x^2-1)(1-y^2)\d\varphi^2\Bigr]\, ,
\label{metric}
\ee
in which the metric coefficients $f$, $\gamma$ and $\omega$ are the
following rational functions of the coordinates $x$ and $y$ (see, e.g.,
\cite{MS} for detais of Sibgatullin's method):
\bea
f&=&E/D, \quad {\rm e}^{2\gamma}=E/16k^8(x^2-y^2)^4, 
\quad \omega=(y^2-1)L/E, \nonumber\\
E&=&\{4[k^2(x^2-1)+\delta (1-y^2)]^2+
(a-b)[(a-b)(d-\delta )-M^2b](1-y^2)^2\}^2
\nonumber\\
&&-16k^2(x^2-1)(1-y^2)\{(a-b)[k^2(x^2-y^2)+2\delta  y^2]+M^2by^2\}^2,
\nonumber\\
D&=&\{4(k^2x^2-\delta  y^2)^2+2kMx[2k^2(x^2-1)+
(2\delta +ab-b^2)(1-y^2)]+(a-b)[(a-b) \nonumber\\
&&\times (d-\delta )-M^2b](y^4-1)-4d^2\}^2+4y^2\{ 2k^2(x^2-1)[kx(a-b)-Mb]
\nonumber\\
&&-2Mb\delta (1-y^2)+[(a-b)(k^2-2\delta )-M^2b](2kx+M)(1-y^2)\}^2,
\nonumber\\
L&=&8k^2(x^2-1)\{(a-b)[k^2(x^2-y^2)+2\delta  y^2]+M^2by^2\} \nonumber\\
&&\times\{ kMx[(2kx+M)^2-a^2+b^2-
2y^2(2\delta +ab-b^2)]-2y^2(4\delta  d-M^2b^2)\}
\nonumber\\
&&-\{4[k^2(x^2-1)+\delta (1-y^2)]^2+(a-b)[(a-b)(d-\delta )-M^2b](1-y^2)^2\}
\nonumber\\
&&\times\Bigl((1-y^2)\{2M(2kx+M)[(a-b)(d-\delta )-b(M^2+2\delta )]
-4M^2b\delta  \nonumber\\
&&+(a-b)(4\delta  d-M^2b^2)\}-8k^2Mb(kx+M)(x^2-1)\Bigr).
\eea

\subsection*{Special cases}

Eqs.~(6) and (10) admit several well-known classical limits:

\begin{enumerate}
\item 
In the absence of the magnetic field and vanishing arbitrary
quadrupole deformation, i.e $\mu=b=0$, only the Ernst potential $\E$
survives which is readily recognizable as that of the Tomimatsu--Sato
$\delta=2$ solution \cite{TS} with the mass quadrupole
$Q=-\frac{1}{4}(M^3+ 3J^2/M)$.

\item 
The stationary pure vacuum limit with a non-vanishing
quadrupole parameter $b$ is a particular three-parameter
specialization of the Kinnersley--Chitre solution \cite{KCh}.

\item
The magnetostatic limit ($a=b=0$) is represented by Bonnor's solution
\cite{B} for a massive magnetic dipole. For this solution, the
quadrupole moment is $Q=\mu^2/M -\frac{1}{4}M^3$. 

\item 
Reduction to the Kerr metric \cite{K} with total mass $M$ and
total angular momentum per unit mass $a:=J/M$ is achieved by setting
$\mu=0$ and then formally choosing $b^2=a^2-M^2$.  The values of $M$
and $a$ remain independent, in particular, $a^2< M^2$ can be imposed
since in this special limit the complex continuation $b\rightarrow \I
b$ is admitted. It should be stressed, however, that there exist
general arguments \cite{He67} for the interior of the Kerr metric
consisting of a perfect fluid according to which a large rotational
flattening of the body necessarily implies a large absolute value 
of the quadrupole moment. 

\item 
It is remarkable that the hyperextreme part of our solution
corresponding to pure imaginary values of $k$ belongs to the
Chen--Guo--Ernst family of hyperextreme spacetimes \cite{ChGE}. This
branch might represent exterior fields of {\em relativistic disks}.
Their importance for astrophysics was shown by Bardeen and Wagoner
\cite{BW}, cf. also \cite{NEL92}. In the absence of the
electromagnetic field an exact global solution for an infinitesimally
thin disk of dust has been constructed by Neugebauer and Meinel
\cite{NM}, cf. \cite{metz}.  
\end{enumerate}

Since neutron stars are known to be `slowly' rotating astrophysical
objects \cite{L} (even at the Kepler frequency $\omega_{\rm
K}=\sqrt{GM/R^3} \simeq 0.5$ ms of a millisecond pulsar would the
equator rotate only with a speed of $v_{\rm K}\simeq c/4$),
therefore, it is the subextreme part of the metric (\ref{metric}) 
which should be used for their description. At the same time, we
still need to know the location of singularities in our solution to
support the physical interpretation we are attributing to it. In
Fig.~1 we have plotted in coordinates $\rho$ and $z$ the typical
shapes of the infinite redshift surface which one has for the
real-valued $k$, the dots indicating the position of singularities.
The two point singularities on the symmetry axis (the poles $x=1$,
$y=\pm1$) belong to the stationary limit surface, while the ring
singularity in each case lies at the equatorial plane between the
symmetry axis and ergosphere; no singularity outside the infinite red
shift surface arises, and hence the metric (\ref{metric}) is
describing the exterior fields of compact objects such as neutron
stars. 

We shall conclude the presentation of our solution by writing out
Kinnersley's potential ${\cal K}$ \cite{Kin} the real part of which
gives the magnetic component $A_\phi$ of the electromagnetic four
potential: 
\bea
{\cal K}&:=&A_\phi + \I A_t^\prime=\mu(1-y^2)K/(A+2MB),\nonumber\\
K&=&2k^2(x^2-1)[2kx+3M+\I y(a-b)]-(a-b)[2Ma-Mb(1-y^2)-4\I\delta  y] 
\nonumber\\
&&+2(2kx+M)[\delta (1-y^2)+M(M-\I by)]+4M\delta \, ,
\eea
where $A_t^\prime$ is the electric component of a vector potential
associated with the dual electromagnetic field tensor. The knowledge
of $A_\phi$ is a necessary basis for the investigation of
plasma-dynamical effects around neutron stars. In Fig.~2 the magnetic
lines of force are plotted in cylindrical coordinates for two
characteristic cases. 

\section{Matching to numerical models of neutron stars}

Our exact axisymmetric solution needs to be {\em matched} to interior 
solutions of neutron stars, in order to be realistic. Since the
junction conditions on the surface of the NS depend very much on the
material details such as equation of state, conductivity etc., we
will restrict ourselves here only to the identification of {\em
asymptotically conserved quantities}. The identification of mass $M$
and angular momentum $J$ of our solution agrees already with the {\em
standard} parameters of asymptotically flat spacetimes in GR
\cite{wini} and in the astrophysics \cite{fried} of NSs.
 
\subsection{Magnetic field}

Normally, the NS' magnetic field $\vec B$, predicted already in 1964
by Hoyle, Narlikar, and Wheeler \cite{[Ho64]} and reaching high
values below the upper limit of $\vec B\leq 3\times 10^9$ Tesla, is
ignored in numerical studies  \cite{fried} of rapidly rotating NS.
More recently, however, axisymmetric solutions of the
Einstein--Maxwell equations have been studied \cite{BBGN95} using a
pseudo-spectral method \cite{Bo93} involving Chebyshev-Legendre
polynomials in terms of maximal slicing quasi-isotropic coordinates. 
Since the magnetic axis is aligned with the rotation axis and only
poloidal fields are permitted, this numerical work is particularly
suited for a comparison of the electrovac spacetime outside the NS
with our exact solution:

For a star close to a sphere and small polar fields $\vec B\sim 10^6$
Tesla, these numerical results are within an error of $10^{-3}$ in
agreement with Ferraro's solution \cite{F54}
\be A_\phi =4\pi\rho_0 f_0 \frac{R_*^5}{15r}\sin^2\theta\, , \qquad
r>R_* \ee 
where $R_*$ is the radius of the star, $\rho_0$ is the mass density,
and $f_0$ is the constant value taken by the electric current
function \cite{BBGN95}.

The  equation above enables us to see how the parameter $\mu$ of our
solution may depend on the parameters of the corresponding interior
metric for small values of $Q$. Indeed, introducing the
Boyer-Lindquist-like coordinates $R$ and $\theta$ via the formulas
\be kx=R-M, \qquad y=\cos\theta \, ,\ee
we easily find from (11), in the limit $R\to\infty$, that 
\be A_\phi=\frac{\mu\sin^2\theta}{R}+O\left(\frac{1}{R^2}\right)
=\frac{\mu\sin^2\theta}{r}+O\left(\frac{1}{r^2}\right)\, , \ee
since $R$ has a representation $R=r(1+O(r^{-1}))$ in terms of the
isotropic coordinate $r$ \cite{fried}. From (12) and (14) the desired
relation of $\mu$ to the parameters determining the interior of a NS
follows immediately.

\subsection{Quadrupole moment}

An operational way of defining the quadrupole moment of an axially
symmetric body in GR has been developed by Ryan \cite{Ry95}. For NS
one finds quite generally \cite{LP99} that the numerical simulations
are rather well accounted by the simple quadratic relation
\be 
Q=-c(M, {\rm EOS})\frac{J^2}{M}\, ,
\label{NSrel}
\ee
where the constant $c=c(M, {\rm EOS})$ depends only on the mass $M$
and the equation of state (EOS) for the interior of the NS. For NSs
of 1.4 $M_\odot$ the range of this constant is $c=2$ to 7.4. It is
intriguing that this simple relation quadratic in $J$ holds also for
fast rotating NS. 
 
If we reparametrize the arbitrary quadrupole parameter $b$ of our
exact solution rather by the dimensionless parameter $\Delta$ via
$b=\pm \sqrt{a^2+2aM\Delta -M^2}$ we obtain from (\ref{quadru}) for
the quadrupole moment ($\mu=0$):  \be Q= -\left(1 +
\frac{1}{2}\Delta^2 - \frac{a\Delta^2(a\Delta^2 -M\Delta \pm
\sqrt{a^2+2aM\Delta -M^2})}{2(M^2 -a^2+a^2\Delta^2
-2aM\Delta)}\right) \frac{J^2}{M} \, .  \ee Our reparametrization is
such that for $\Delta =0$ we recover the mass-quadrupole moment
$Q=-J^2/M$ of the Kerr metric.  Thus it is worth pointing out that
the mass-quadrupole parameter in our solution is also intimately
related with the angular momentum dipole and octupole moments, and
this means that the {\em deformations} of the source are mainly due
to {\em rotation}. In a particular case, for instance, when
$M=1.4M_\odot$ and $a=0.625M_\odot$, the values of $\Delta$ covering
the NS range $2<c<7.4$ are given by the interval
$0.986<\Delta<2.068$.

In comparison with (\ref{NSrel}), the constant $c$ for our solution
depends not only on the mass $M$, but also on $\Delta >(M^2-a^2)/2aM$
which can be adjusted to different EOS. The additional piece
depending on angular momentum per unit mass, however, arises only in
higher order of $\Delta$. Thus the quadrupole moment of our exact
solution accounts rather well to the simple quadratic law
(\ref{NSrel}) of NSs. Moreover, in the `extreme' limit $a \rightarrow
M$ we find
\be 
Q=-\left(1 -
\frac{\Delta(\Delta 
\pm \sqrt{2\Delta})}{2(\Delta 
-2)}\right) M^3\, ,
\ee
whence it can be seen that $Q$ can assume arbitrary values for any
given value of $M$, unlike in the case of the extreme Kerr metric for
which $Q=-M^3$. In this particular limit, the `NS interval'
corresponds to $0.931<\Delta<1.584$.

Futher study is needed to exhibit the astrophysical significance of
our solution in more detail. In view of a simple analytical form of
the new electrovacuum metric and clear physical interpretation of the
characteristic parameters it possesses, it is anticipated that the
solution will prove itself suitable for the use in concrete
astrophysical applications involving neutron stars. \vspace{0.8cm}

\section*{Acknowledgments}

It is our pleasure to thank Jerzy Pleba\'nski and Alfredo Mac\'\i as
for stimulating discussions and support. We are grateful to
Olga~Manko for her help in establishing the factor structure of the
metric coefficients (10). This work was partially supported by
Conacyt, grant No.~28339--E, as well as the joint German--Mexican
project DLR--Conacyt E130--1148 and MXI 010/98 OTH. One of us
(J.D.S-G.) acknowledges financial support from Colciencias of
Colombia.

\begin{figure}[htb]
\centerline{\epsfysize=8cm\epsfbox{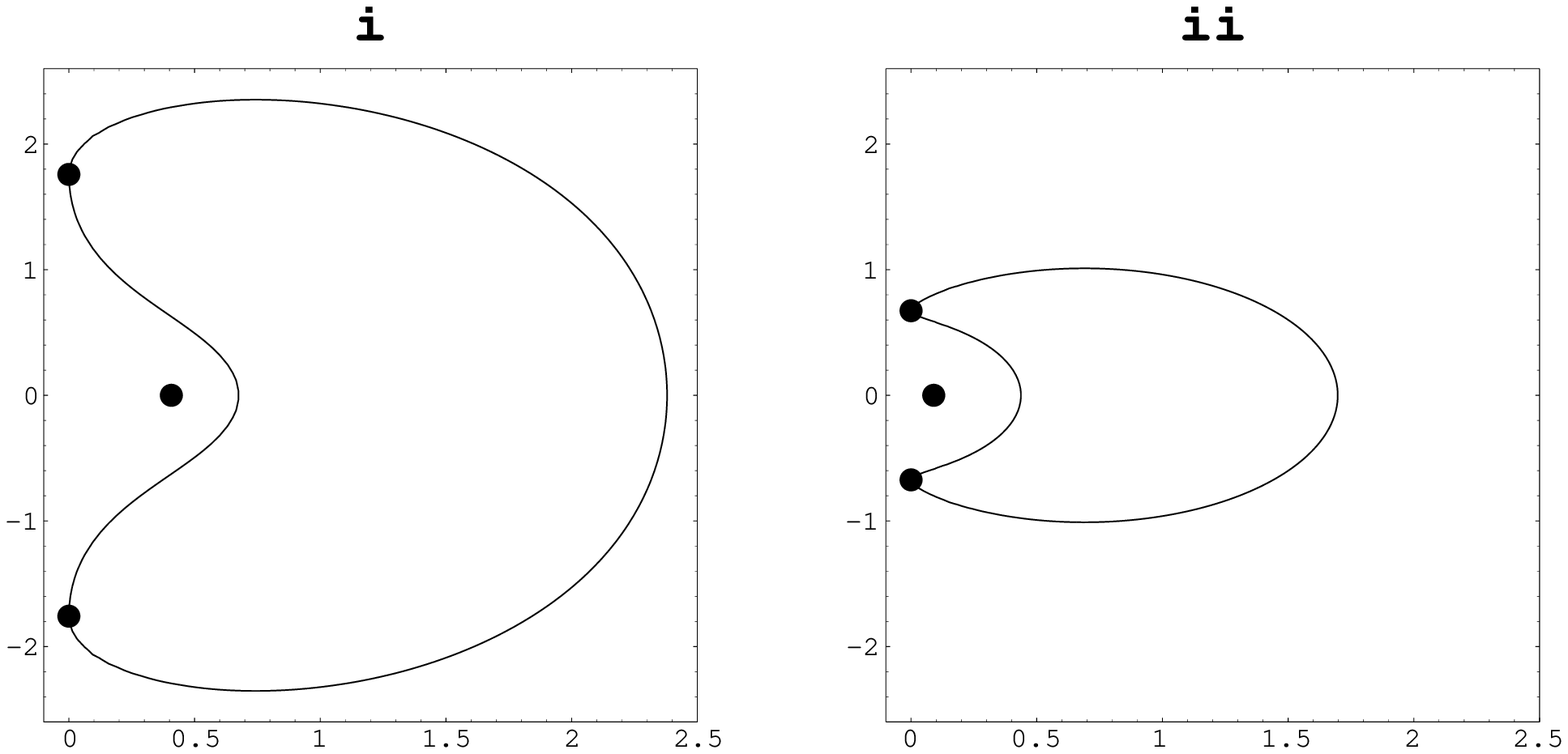}}
\caption{Ergosphere and singularities:  i) M = 4, a = 2, b = 0.9,
$\mu$ = 2 ;
ii) M = 2, a = 1.6, b = -0.2, $\mu$ = 0.6 (m or m$^2$, in the natural
units).}
\end{figure}

\begin{figure}[htb]
\centerline{\epsfysize=8cm\epsfbox{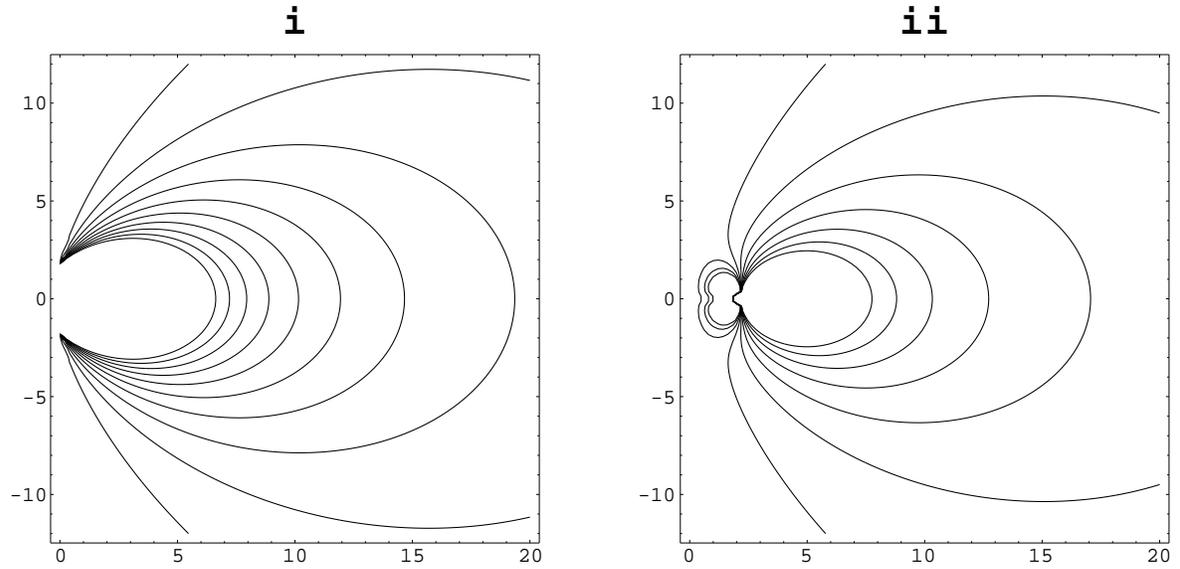}}
\caption{Magnetic lines of force:  i) M = 4, a = 2, b = 0.4,
$\mu$ = 1.2;
ii) M = 2, a = 4, b = -1, $\mu$ = 3 (m or m$^2$, in the natural
units).}
\end{figure}


\newpage

\begin{references}
\bibitem{TAMT93} S.~E.~Thorsett, Z.~Arzoumanian, M.~M.~McKinnon and
J.~Taylor, \J{\AJ}{405}{L29}{1993}.
\bibitem{BB94} 
G.~E.~Brown and H.~Bethe, \J{\AJ}{423}{659}{1994}.
\bibitem{GHL99} E.~Gourgoulhon, P.~Haensel, R.~Livine, E.~Paluch,
S.~Bonazzola and J.-A.~Marck, \J{\AA}{349}{851}{1999}.
\bibitem{BSS99} Th.~W.~Baumgarte, S.~L.~Shapiro and M.~Shibata,
\J{\AJ}{528}{L29}{2000}.
\bibitem{SS} N.~R.~Sibgatullin, and R.~A.~Sunyaev, \J{\AL}{24}{894}{1998}.
\bibitem{C} G.~B.~Cook, S.~L.~Shapiro, and S.~A.~Teukolsky,
\J{\AJ} {422}{227}{1994}.
\bibitem{LP99} W.~G.~Laarakkers and E.~Poissson,
\J{\AJ}{512}{282}{1999}. 
\bibitem{MN} R.~Meinel, and G.~Neugebauer, \J{\CQG}{12}{2045}{1995}.
\bibitem{A-S} J.~M.~Aguirregabiria, A.~Chamorro, V.~S.~Manko, and
N.~R.~Sibgatullin, \J{\PRD}{48}{622}{1993};
\bibitem{M-Z} V.~S.~Manko, J.~Mart\'\i n, E.~Ruiz, N.~R.~Sibgatullin,
and M.~N.~Zaripov, \J{\PRD}{49}{5144}{1994};
V.~S.~Manko, J.~Mart\'\i n, E.~Ruiz, \J{\JMP}{36}{3063}{1995}.
\bibitem{S} N.~R.~Sibgatullin: {\em Oscillations and Waves in Strong
Gravitational and Electromagnetic Fields} (Nauka, Moscow, 1984; English
translation: Springer-Verlag, Berlin, 1991). 
\bibitem{E} F.~J.~Ernst, \J{\PR}{168}{1415}{1968}.
\bibitem{W} S.~Wolfram: {\em Mathematica}  (Addison-Wesley Publishing
Company, 1991).
\bibitem{MS} V.~S.~Manko and N.~R.~Sibgatullin,
\J{\CQG}{10}{1383}{1993}.
\bibitem{TS} A.~Tomimatsu, and H.~Sato, \J{\PRL}{29}{1344}{1972};
\J{\PTP}{50}{95}{1973}.
\bibitem{KCh} W.~Kinnersley, and D.~M.~Chitre, \J{\JMP}{19}{2037}{1978}.
\bibitem{B} W.~B.~Bonnor, \J{\ZP}{190}{444}{1966}.
\bibitem{K} R.~P.~Kerr, \J{\PRL}{11}{237}{1963}.
\bibitem{He67} W.~C.~Hernandez, Jr., \J{\PR}{159}{1070}{1967}.
\bibitem{ChGE} Y.~Chen, D.~S.~Guo, and F.~J.~Ernst 
\J{\JMP}{24}{1564}{1983}.
\bibitem{BW} J.~M.~Bardeen, and R.~V.~Wagoner, \J{\AJ}{158}{L65}{1969};
\J{\AJ}{167}{359}{1971}.
\bibitem{NEL92} S.~Nishida, Y.~Eriguchi and A.~Lanza, \J{\AJ}
{401}{618}{1992}.
\bibitem{NM} G.~Neugebauer, and R.~Meinel, \J{\AJ}{414}{L97}{1993};
\J{\PRL}{73}{2166}{1994}; \J{\PRL}{75}{3046}{1995}.
\bibitem{metz} R.~Metzler, \J{\PLA}{210}{45}{1996}.
\bibitem{L} J.~D.~Landstreet, \J{\AAR}{4}{35}{1992}.
\bibitem{Kin} W.~Kinnersley, \J{\JMP}{18}{1529}{1977}.
\bibitem{wini}
J.~Winicour, in: {\em General Relativity and Gravitation --- One
Hundred Years After the Birth of Albert Einstein (Vol.~2)},
ed.~A.~Held (Plenum Press, New York 1980), pp.~71-96.
\bibitem{fried}
J.~L.~Friedman and J.~R.~Ipser,  Phil. Trans. R. Soc. Lond. A
{\bf 340}, 391 (1992).
\bibitem{[Ho64]} F.~Hoyle, J.~V.~Narlikar and J.~A.~Wheeler, Nature
{\bf 203}, 914 (1964). 
\bibitem{BBGN95} M.~Bocquet, S.~Bonazzola, E.~Gourgoulhon and
J.~Novak, \J{\AA}{301}{757}{1995}.
\bibitem{Bo93}
S.~Bonazzola, E.~Gourgoulhon, M.~Salgado, and J.~A.~Marck, 
\J{\AA}{278}{421}{1993}.
\bibitem{F54} V.~C.~A.~Ferraro, \J{\AJ}{118}{407}{1954}.
\bibitem{Ry95} F.~D.~Ryan, \J{\PRD}{52}{5707}{1995}.

\end{references}
\end{document}